\documentclass[twocolumn]{aastex62}
\usepackage{graphics,epsf}
\usepackage{amsmath}                
\usepackage{amsfonts}               
\usepackage{amssymb}                
\usepackage{epsfig}                 
\usepackage{graphicx}
\usepackage{float}
\usepackage{color}
\usepackage[para,online,flushleft]{threeparttable}

\newcommand{\K}{{~\rm K}}

\newcommand{\yr}{{~\rm yr}}

\newcommand{\AU}{{~\rm AU}}

\begin{document}

\title{Eccentric grazing envelope evolution towards 
type~IIb supernova progenitors} 


\author{Dmitry Shishkin}
\affiliation{Department of Physics, Technion, Haifa, 3200003, Israel; s.dmitry@campus.technion.ac.il; soker@physics.technion.ac.il}

\author[0000-0003-0375-8987]{Noam Soker}
\affiliation{Department of Physics, Technion, Haifa, 3200003, Israel; s.dmitry@campus.technion.ac.il; soker@physics.technion.ac.il}
\affiliation{Guangdong Technion Israel Institute of Technology, Shantou 515069, Guangdong Province, China}

\begin{abstract}
We simulate the evolution of eccentric binary systems in the frame of the grazing envelope evolution (GEE) channel for the formation of Type~IIb supernovae (SNe~IIb), and find that extra mass removal by jets increases the parameter space for the formation of SNe~IIb in this channel.
To explore the role of eccentricity and the extra mass removal by jets we use the stellar evolutionary code \textsc{MESA~binary}. The initial primary and secondary masses are $M_{\rm 1,i}=15M_\odot$ and $M_{\rm 2,i}=2.5M_\odot$. We examine initial semi-major axes of $600-1000 R_\odot$, and eccentricities of $e=0 - 0.9$.
Both Roche lobe overflow (RLOF) and mass removal by jets, followed by a wind, leave a hydrogen mass in the exploding star of $M_{\rm H,f} \approx 0.05 M_\odot$, compatible with a SN~IIb progenitor. 
{{{{ When the initial orbit is not circular the final orbit might have a very high eccentricity. }}}} 
In many cases, with and without the extra mass removal by jets, the system can enter a common envelope evolution (CEE) phase, and then gets out from it. 
{{{{  For some ranges of eccentricities the jets are more efficient in preventing he CEE.  }}}}
Despite the large uncertainties, extra mass removal by jets substantially increases the likelihood of the system to get out from a CEE. This strengthens earlier conclusions for circular orbits. 
In some cases RLOF alone, without mass removal by jets, can form SN IIb progenitors.
We estimate that the extra mass removal by jets in the GEE channel increases the number of progenitors relative to that by RLOF alone by about a factor of two. 
\end{abstract} 


\section{INTRODUCTION}
\label{sec:intro}

Core collapse supernovae (CCSNe) of type IIb (SNe~IIb) show strong hydrogen lines days after explosion, but only weak hydrogen lines, or none at all, at later times. This large weakening of the hydrogen lines results from a very small hydrogen mass in the envelope of the exploding star, namely, $M_{\rm H} \simeq 0.03-0.5 M_\odot$ (e.g., \citealt{Woosleyetal1994, Meynetetal2015, Yoonetal2017}), or $0.01 M_\odot \le M_{\rm H,env} \le 1 M_\odot$ (e.g., \citealt{Sravanetal2018}), or even down to $ M_{\rm H} \simeq 0.001 M_\odot$ \citep{Dessartetal2011, Eldridgeetal2018}. A fraction of $f_{\rm IIb} \simeq 11 \%$ of all CCSNe are SNe~IIb \citep{Smithetal2011, Shivversetal2017, Grauretal2017b, Sravanetal2018}, with up to $f_{\rm IIb, L} \simeq 20 \%$ in low metallicity populations \citep{Sravanetal2018}.

One possible classification of SNe~IIb progenitors is to compact progenitors, i.e., blue progenitors and yellow supergiant progenitors that lead to most SNe~IIb (e.g., \citealt{Yoonetal2017}), and to extended progenitors, i.e., red supergiants (e.g., \citealt{ChevalierSoderberg2010}).  The compact progenitors have small hydrogen mass, $M_{\rm H} \la 0.15 M_\odot$, at explosion, 
while the red supergiant progenitors have $M_{\rm H} \ga 0.15M_\odot$ at explosion \citep{Yoonetal2017}. Mass transfer, both stable and unstable, followed by winds that are efficient in removing most of the remaining hydrogen can form compact SN IIb progenitors (e.g., \citealt{Yoonetal2017, Gilkisetal2019}). 
Winds in higher metallicity populations are more efficient in removing mass, therefore leading to a higher ratio of SNe~Ib to SNe~IIb (e.g., \citealt{Yoonetal2017}). 

Several studies of specific SNe~IIb attribute the low hydrogen mass at explosion to binary interaction, e.g., SN~1993J \citep{Podsiadlowskietal1993, Alderingetal1994, Maundetal2004, Foxetal2014}, SN~2016gkg; \citep{Kilpatricketal2017, Berstenetal2018},  and  ZTF18aalrxas \citep{Fremlingetal2019}. 
Other studies examine the general population of SNe IIb (e.g., 
\citealt{StancliffeEldridge2009, Claeys2011, OuchiMaeda2017}), concluding also that to account for a large fraction of SNe IIb the binary system progenitors should lose mass more efficiently than what traditional binary evolution predict. 

\cite{Sravanetal2018} consider Roche lobe overflow (RLOF) mass transfer in their study of SN IIb progenitors, and conclude from their population synthesis study that the binary channel and the single stellar channel have about equal contribution to SNe IIb (also \citealt{Sravan2016}). Their study, however, falls short by more than a factor of three in accounting for the rate of SNe IIb. Weaker winds than what traditional mass loss rate formulae give might leave some hydrogen in the envelope after the end of the mass transfer process; this eases the tension with observations \citep{Gilkisetal2019}. 

There are two more binary channels in addition to that of the RLOF. The second evolutionary channel involves a common envelope evolution (CEE) with a main sequence companion that enters the giant envelope \citep{Nomotoetal1995, Youngetal2006, Lohevetal2019}. \cite{Lohevetal2019} consider a main sequence companion that spirals-in inside the envelope, ejects the entire original hydrogen-rich envelope, and reaches the core. The core tidally destroys the companion (a fatal-CEE) such that the companion material forms a new low-mass hydrogen-rich envelope of the massive star. The star explodes later as a SN IIb, as   \cite{Lohevetal2019} suggest for SN~IIb Cassiopeia~A (see, e.g., \citealt{Satoetal2020} for a recent paper with arguments for a binary model for Cassiopeia~A).  
   
The third binary channel involves the grazing envelope evolution (GEE). \cite{Soker2017} proposes that in some cases the extra mass removal that some studies require to form SNe IIb (e.g., \citealt{Claeys2011, OuchiMaeda2017}) can result from jets that the companion to the exploding star launches. The binary companion launches the jets as it grazes the envelope of the more massive progenitor of the SN IIb. The jets remove mass efficiently from the envelope, and by that postpone or prevent the onset of a CEE; instead, the binary system experiences the grazing envelope evolution (GEE; \citealt{Soker2015}).  {{{{ Recent three-dimensional hydrodynamical simulations show indeed that jets that the companion launches can enhance mass removal during the CEE and GEE (e.g., \citealt{ShiberSoker2018, LopezCamaraetal2019, Shiberetal2019, LopezCamaraetal2020}). }}}}
\cite{Naimanetal2020} study the GEE channel for circular binary orbits with the \textsc{binary} module of the \textsc{mesa} code (Modules for Experiments in Stellar Astrophysics;  \citealt{Paxtonetal2011, Paxtonetal2019}) by mimicking the extra mass loss due to jets with a simple numerical prescription. The GEE channel forms mainly blue-compact SN~IIb progenitors, because post-GEE winds remove most, but not always  all, of the remaining hydrogen. 
   
Overall, there are four channels for SN IIb progenitors, i.e., exploding star with little hydrogen mass. These are the single star evolution, RLOF binary evolution, the fatal-CEE, and the GEE.   
\cite{Naimanetal2020} crudely estimate that each of the four channels contributes about equally (25 per cents) to the SN IIb population.  \cite{Naimanetal2020} present more details on the relevant parts of the GEE channel, on the general motivation to consider the GEE, and discuss the qualitative differences between the GEE and the RLOF channels. One major difference is that in the RLOF process the companion orbits well outside the giant envelope, while in the GEE the companion grazes the giant envelope. In the present study we examine eccentric orbits such that the companion grazes the giant at and near periastron passages. 
 
To further support the GEE scenario, in this study we extend the study of  \cite{Naimanetal2020} and include GEE with eccentric orbits (section \ref{sec:Mimicking}). The new results add to the rich variety of possible outcomes of the GEE, including the increase of the parameter space for SN IIb formation (sections \ref{sec:eccentricity} and \ref{sec:OtherCases}), and strengthen the GEE channel for the formation of some SN IIb progenitors. We summarise in section \ref{sec:summary}.  
 
\section{MIMICKING THE GRAZING ENVELOPE EVOLUTION}
\label{sec:Mimicking}
\subsection{Binary evolution}
\label{subsec:Binary}

We conduct binary evolution simulations with the binary module of the \textsc{mesa} code (version 10398; \citealt{Paxtonetal2011, Paxtonetal2013, Paxtonetal2015, Paxtonetal2018, Paxtonetal2019}), with the goal of demonstrating that the GEE in eccentric orbits can increase the parameter space for the formation of SNe~IIb progenitors. At this stage we limit our study to a small number of cases to explore the properties of this evolutionary channel. Due to some uncertainties, like the simple prescription we use to mimic the effects of jets and uncertainties in the wind mass loss rate (e.g, \citealt{Gilkisetal2019, Beasoretal2020}), we are not yet in a position to directly derive the fraction of SNe~IIb that result from the GEE channel. This is only the third set of GEE calculations with \textsc{mesa~binary}, and each study has different parameters \citep{AbuBackeretal2018, Naimanetal2020}. 

Key assumptions of the GEE study with \textsc{mesa~binary} are  as follows. (1) The specific angular momentum of the mass that the primary (giant) star transfers to the secondary star is sufficient to form an accretion disk around the secondary star. (2) The accretion disk launches jets, as accretion disks around young stellar objects do. (3) When the secondary star is very close to the surface of the giant, or even somewhat inside the envelope,  jets efficiently remove mass from the outer envelope.

At zero age main sequence the binary systems we simulate have the following properties. The initial mass of the primary (mass donor star) is $M_{\rm 1,i}=15 M_\odot$ and its metallicity is $Z=0.019$. The initial mass of the secondary star is  $M_{\rm 2,i}=2.5 M_\odot$.
We set the code to treat the secondary star as a point mass, i.e., we do not follow its evolution (see \citealt{Naimanetal2020}). We set the initial semi-major axis to be in the range of $a=600-1000 R_\odot$, and the eccentricity to be in the range of $e=0$ to $e=0.9$ (note that \citealt{Naimanetal2020} considered only $e=0$). 

The \textsc{binary} module of \textsc{mesa} evolves the system according to the stellar evolution of the primary star, mass transfer, mass loss, and tidal interaction that changes both the eccentricity and the semi-major axis of the orbit (\citealt{Hut1981}, with the timescales for convective envelopes from \citealt{Hurley2002}), and the spin of the primary star. The \textsc{mesa} code sets rotation according to the `shellular approximation', with constant angular velocity $\omega$ on isobars (e.g., \citealt{Meynet1997}). 

In some cases we let the system to evolve till core collapse even if periastron brings the companion into the primary envelope. In other cases we terminate the evolution when the secondary enters the envelope,  Namely, when 
\begin{equation}
a(1-e) < R_1+R_2.  
\label{eq:aR1R2}
\end{equation}
We take $R_2=0$ in all cases.
We terminate the evolution in some cases because when the secondary star is inside the giant envelope the simple formulae for tidal interaction are not accurate anymore. More over, we have larger uncertainties concerning the launching of jets (see also \citealt{Naimanetal2020}). 

To learn about the role of the jets, we also simulate cases without jets, and compare the two types of simulations. 
In all cases the RLOF mass transfer rate $\dot M_{\rm KR}$, is from  \cite{Kolb1990}, {{{ while the Roche-lobe radius is from \cite{Eggleton1983}. }}} In all simulations we take a fraction $f_{\rm acc,RL}=0.3$ of the RLOF mass transfer to be accreted by the secondary star. The rest of the mass that the primary transfers is lost by the secondary star, i.e., a fraction of $f_{\rm L,RL,2}=0.7$.
We do not change all variables, but rather keep many of them the same in the different runs, because in the present study we are interested in the role of the eccentricity in the GEE channel. 

\subsection{Mimicking jets}
\label{subsec:jets}

We follow \cite{Naimanetal2020} in mimicking the extra mass removal by jets. 
According to the GEE the jets that the secondary star launches remove mass from the outer parts of the envelope of the primary star, and from the acceleration zone of its wind \citep{Hilleletal2020}. We set mass removal by jets to take place when the following condition applies    
\begin{equation}
a (1-e)<f_\mathrm{GEE} \left(R_1+R_2\right),  
\label{eq:fGEE}
\end{equation}
where $R_2=0$ in the present study. The jet-activity separation factor takes the values of $f_\mathrm{GEE}=1.1$ or $f_\mathrm{GEE}=1.2$. 
When the inequality of equation (\ref{eq:fGEE}) holds, we take an additional mass loss from the system, half of it from the primary star and half of it from the secondary star. The total jet-driven mass loss rate is \begin{equation}
\dot M_{\rm L,jet} = f_{\rm jet} \dot M_{\rm KR}  
\frac{f_\mathrm{GEE} - a(1-e)/R_1} {f_\mathrm{GEE} - 1} ,
\label{eq:MLJ1}
\end{equation}
where we introduced the jet-driven mass loss factor $f_{\rm jet}$.  
In the present study we take $f_{\rm jet}=3$ as \cite{Naimanetal2020} suggest, and $f_{\rm jet}=4$ for some runs (Table \ref{Table2}). 

When the activity of the jets begins we reduce the time step by setting the \textsc{mesa} variable \texttt{varcontrol\char`_target} to $10^{-5}$ instead of the default value of $10^{-4}$.

\subsection{Wind mass loss rate}
\label{subsec:wind}

After the end of the binary mass transfer, the wind from the primary star removes substantial amounts of mass. There are large uncertainties in wind mass loss rates from red supergiants and other luminous stars (e.g., \citealt{Gilkisetal2019}). \cite{Beasoretal2020}, for example, argue that models overpredict the total mass-loss by a large factor. For that, the mass loss rate might be lower even than what we use here. 

We proceed as in \cite{Naimanetal2020}. The wind mass loss rate is according to \cite{deJager1988} with scaling factor of 1 for an effective temperature of $T_\mathrm{eff} < 10^4 \K$. For $T_\mathrm{eff} \ge 1.1 \times 10^4 \K$ we follow the mass loss rate from \cite{Vink2001} if the surface hydrogen mass fraction is  $X_\mathrm{s} > 0.4$ or from \cite{Vink2017} when $X_\mathrm{s} \le 0.4$. We interpolate for $1.1\times 10^4\,\mathrm{K}>T_\mathrm{eff} > 10^4\,\mathrm{K}$. 

\subsection{Numerical limitations}
\label{subsec:Numerical}

\cite{Naimanetal2020} discuss in length the numerical limitations of mimicking the GEE in the \textsc{mesa~binary} code. We briefly summarise their discussion. 
The two main limitations are the poor handling of the CEE, and the large number of free parameters, e.g., the mass loss scheme.
Other free parameters include the form of equation (\ref{eq:MLJ1}) and the parameter $f_{\rm jet}$ there, the parameter $f_{\rm GEE}$ in equation \ref{eq:fGEE}, the parameters of the RLOF (section \ref{subsec:Binary}), and the fraction of mass that each star loses during the jets activity episodes (in this study each star loses half of the mass; section \ref{subsec:jets}). 

The CEE is very complicated, and more so when the secondary star launches jets. In some simulations, with and without jets, according to \textsc{mesa~binary} the binary system gets into a CEE phase, and then out (\citealt{Naimanetal2020}, and in section \ref{sec:eccentricity}). However, this evolutionary track of in-and-out of the CEE is highly uncertain and not accurate.   
Basically, MESA does not know it enters a CEE phase. It just continues to compute the mass loss according to the RLOF formulae, as by definition the primary overfills its Roche lobe during CEE.
The extra mass loss by jets that we use here is proportional to the RLOF mass transfer rate (equation \ref{eq:MLJ1}), and it continues  inside the envelope. The results after a long CEE phase should be treated with a high cautious. 

Without jets, we expect that in a circular orbit when the secondary enters the giant envelope it will continue to spiral-in. In orbits with high eccentricity the secondary might get out without jets. Jets might ease the exit from a CEE, and the system continues with the GEE. Again, these transitions from the GEE to the CEE and back, are highly uncertain. 
 
New to this study is the limitation in our implementation of the jets mechanism for eccentric orbits. Specifically, in equation (\ref{eq:fGEE}) we turn the jets on according to the ratio of the periastron distance to the primary radius.
If the condition holds, jets are active along the entire orbit. 
We do not follow the orbital separation along the orbit. Such a more accurate treatment requires a different and a much more extended study that will introduce more free parameters.  
In any case, we do not think this is a major limitation because the RLOF mass transfer scheme in \textsc{mesa~binary} takes eccentricity into account when averaging the mass transfer over one orbit. Because our mass removal by jets scheme is proportional to the RLOF mass transfer rate (equation \ref{eq:MLJ1}), we also take this into account. 
We also found that when we start with $e_i \ga 0.8$ the eccentricities at late times might become close to 1, when we cannot trust the evolutionary scheme for our purposes. 

\section{The role of eccentricity}
\label{sec:eccentricity}

\subsection{Cases with circular orbits}
\label{subsec:circOrbits}

\cite{Naimanetal2020} study the cases with circular orbits, i.e., the eccentricity is forced to be $e=0$ while the semi-major axis $a$ changes along the evolution. We will not repeat their results, but rather present only one case, with and without jets, for later comparisons with eccentric orbits. 
We present in Fig. \ref{fig:1000,e=0,fiducialded wind} the results for an initial orbital radius of $a_0=1000 R_\odot$, and component masses of $M_{\rm 1,i}=15 M_\odot$ and $M_{\rm 2,i}=2.5 M_\odot$. The jets-induced enhanced mass loss rate parameters are $f_{\rm GEE}=1.1$ (equation \ref{eq:fGEE}) and $f_{\rm jet} = 3$ (equation \ref{eq:MLJ1}). 
The left panels present the case for which the jets become active when condition (\ref{eq:fGEE}) holds, while the right panels present the case where we do not allow for jets activity at all.  
Black lines in the upper panels present the orbital radius $a$, the thick solid blue lines represent the radius of the primary star $R_1$, and the dashed-blue lines present the distance $(1-e)a/1.1$ (which here is $a/1.1$) that appears in condition (\ref{eq:fGEE}) for triggering jets activity. The dashed-red lines represent the Roche-lobe radius. 
The lower panels present the masses of the primary $M_1$ (thick blue lines) and secondary $M_2$ (thick red lines), and the hydrogen mass in the primary star $M_{\rm H,1}$ (thin black lines).
\begin{figure*}
\begin{center}
 \includegraphics[trim=2cm 2cm 2cm 3cm,scale=0.65]{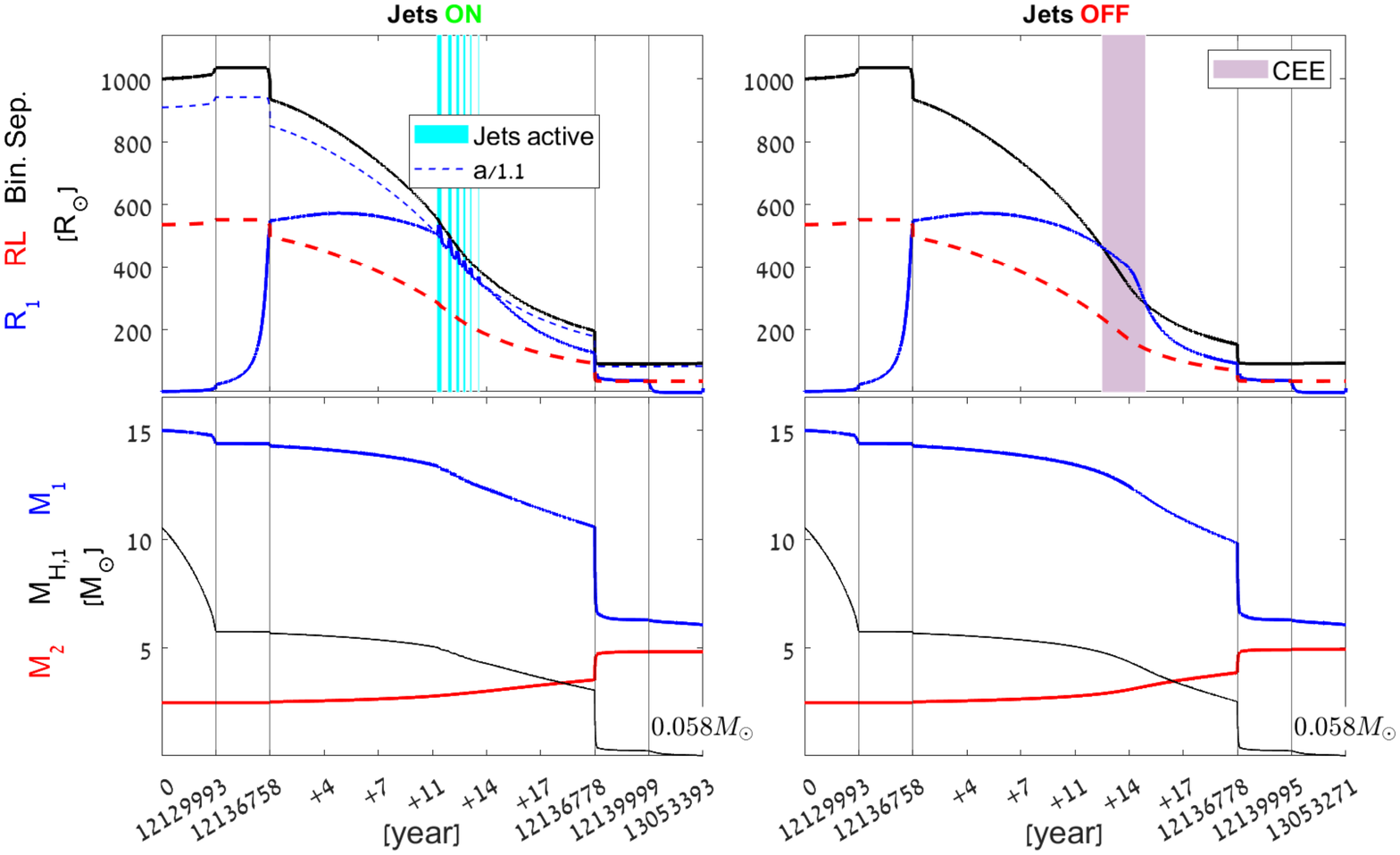}
\caption{Evolution for circular orbits and an initial orbital separation of $a_0=1000 R_\odot$, masses of $M_{\rm 1,i}=15 M_\odot$ and $M_{\rm 2,i}=2.5 M_\odot$, with (left panels) and without (right panels) jets activity. 
The panels are separated (thin vertical lines) into five time periods with different scaling, as we mark on the horizontal axes in years. The five time periods are, from left to right, $1.213 \times 10^7 \yr$ (earliest time segment), $6765 \yr$, $20 \yr$, $3221/3217 \yr$, and $9.314 \times 10^5 \yr$ (last time segment) for the left/right panels.
Upper panels: orbital radius $a$ (black lines), the radius of the primary star $R_1$ (thick solid blue lines), the distance $(1-e)a/1.1$ (which here is $a/1.1$; dashed-blue lines) that appears in condition (\ref{eq:fGEE}) for triggering jets activity, and the Roche-lobe radius (dashed-red lines). 
Lower panels: primary mass $M_1$ (thick blue lines), secondary mass $M_2$ (thick red lines), and the hydrogen mass $M_{\rm H,1}$ (thin black lines).  
Teal bands on the left upper panel denotes when jets are active, and the lilac band on the right upper panel denotes a CEE phase.
Both cases end with a hydrogen mass of $0.058 M_\odot$ at explosion, forming SNe IIb. }
\label{fig:1000,e=0,fiducialded wind}
\end{center}
\end{figure*}
 
The large difference between the two cases is that the evolution with no jets activity enters a CEE (according to equation \ref{eq:aR1R2}), while when we allow jets activity the system avoids the CEE. We mark the time period when the system enters the CEE with a lilac vertical band on the upper right panel. The system then exists the CEE, and ends as a SN IIb progenitor. When we allow jets activity, we find that the binary system experiences six jets activity periods, which we mark by teal vertical bands on the left upper panel of Fig. \ref{fig:1000,e=0,fiducialded wind}. 
Both cases end as SN IIb progenitors with a hydrogen mass of $M_{\rm H,f}=0.058 M_\odot$, same as in \cite{Naimanetal2020}, where they give more details on circular orbits. In particular, they emphasise that there are large uncertainties in the evolution when the system enters a CEE, and it is possible that in reality the binary system will not exit the CEE in the case without jets activity (section \ref{subsec:Numerical}). Indeed, 
there are some cases when the binary system do not exit the CEE. 

Overall, \cite{Naimanetal2020} conclude from their study of circular orbits that allowing jets activity increases the parameter space for the formation of SN IIb progenitors. We now turn to examine the role of eccentricity. 

\subsection{Eccentric orbits}
\label{subsec:EccentricOrbits}

We present now cases with eccentric orbits, keeping all other initial parameters as in the circular cases of section \ref{subsec:circOrbits} (Fig. \ref{fig:1000,e=0,fiducialded wind}). 

In Fig. \ref{fig:1000,e=0.1,fiducialded wind} we present the evolution for an initial eccentricity of $e_i=0.1$, with allowed jet activity (according to condition \ref{eq:fGEE}; left panels) and without jet activity (right panels). In this case there is no jets activity at all, and the system avoids the CEE. Both cases therefore, are the same. We attribute this behavior to the role of tidal interaction that is stronger at periastron, and as a result of that the primary star in the $e_i=0.1$ case does not reach a large maximum radius as in the circular case (compare thick blue lines in the upper panels of both figures). {{{{ The large increase in eccentricity that occurs during the mass transfer results from more efficient mass loss near periastron passages (e.g., \citealt{KashiSoker2018}). }}}}
\begin{figure*}
\begin{center}
 \includegraphics[trim=2cm 2cm 2cm 3cm,scale=0.65]{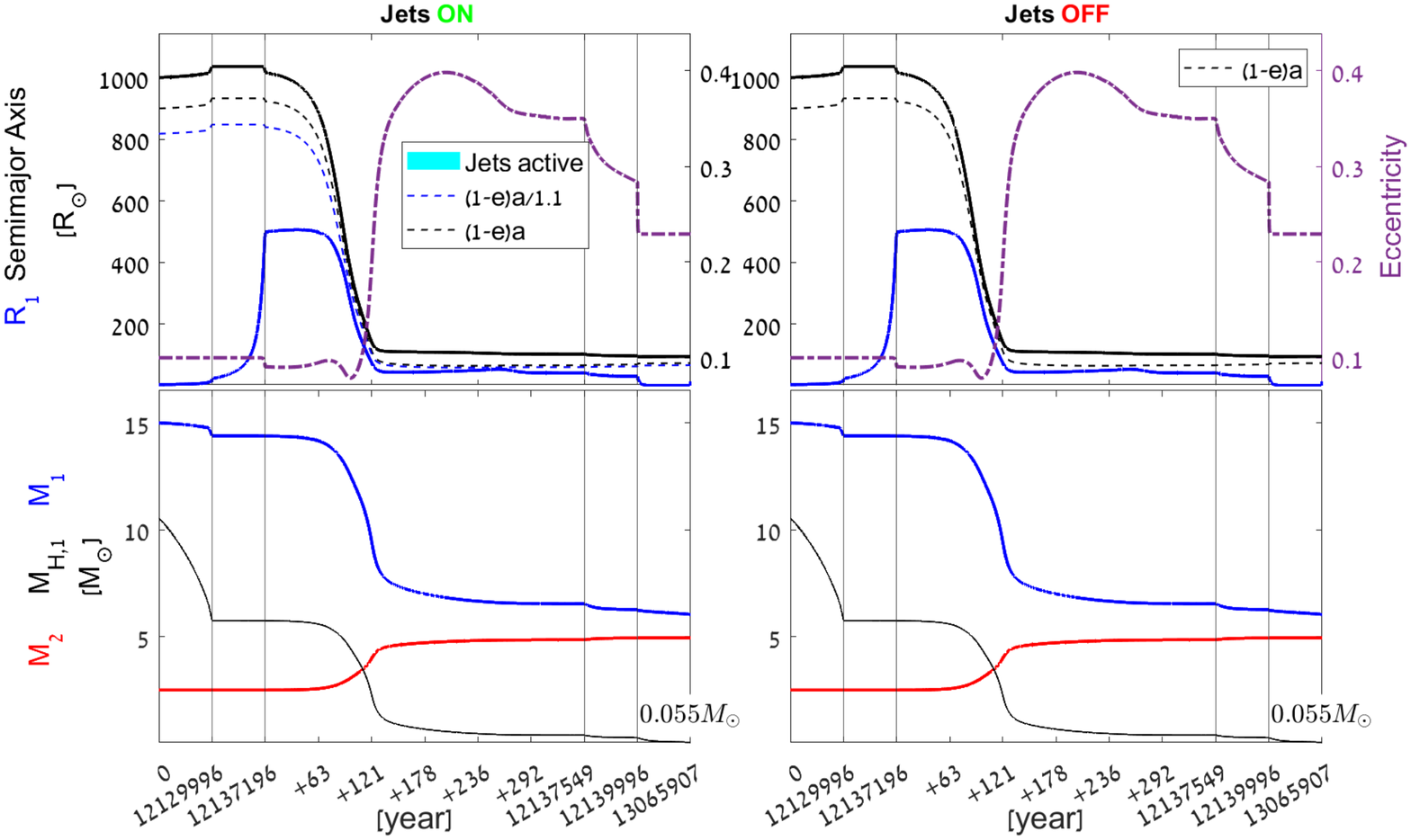}
\caption{Like Figure \ref{fig:1000,e=0,fiducialded wind}, but for an initial eccentricity of $e_i=0.1$ rather than for a circular orbit, and with the presentation of two more parameters. 
The dashed-doted purple lines in the upper panels depict the evolution of the eccentricity, with scale on the right axis. The dashed-black line in the upper panel is the periastron distance $(1-e)a$. 
The five time periods span, from left to right, $1.213 \times 10^7 \yr$ (earliest time segment), $7200 \yr$, $353 \yr$, $2447 \yr$, and $9.259 \times 10^5 \yr$ (last time segment). 
In contrast with the circular orbit, here even without jets activity the system avoids the CEE, and the jets never turned on. Therefore, the panels on both sides of the figure are identical. 
}
\label{fig:1000,e=0.1,fiducialded wind}
\end{center}
\end{figure*}
 
\begin{figure}[H]
\begin{center}
\includegraphics[trim=2cm 5cm 2cm 5cm,scale=0.4]{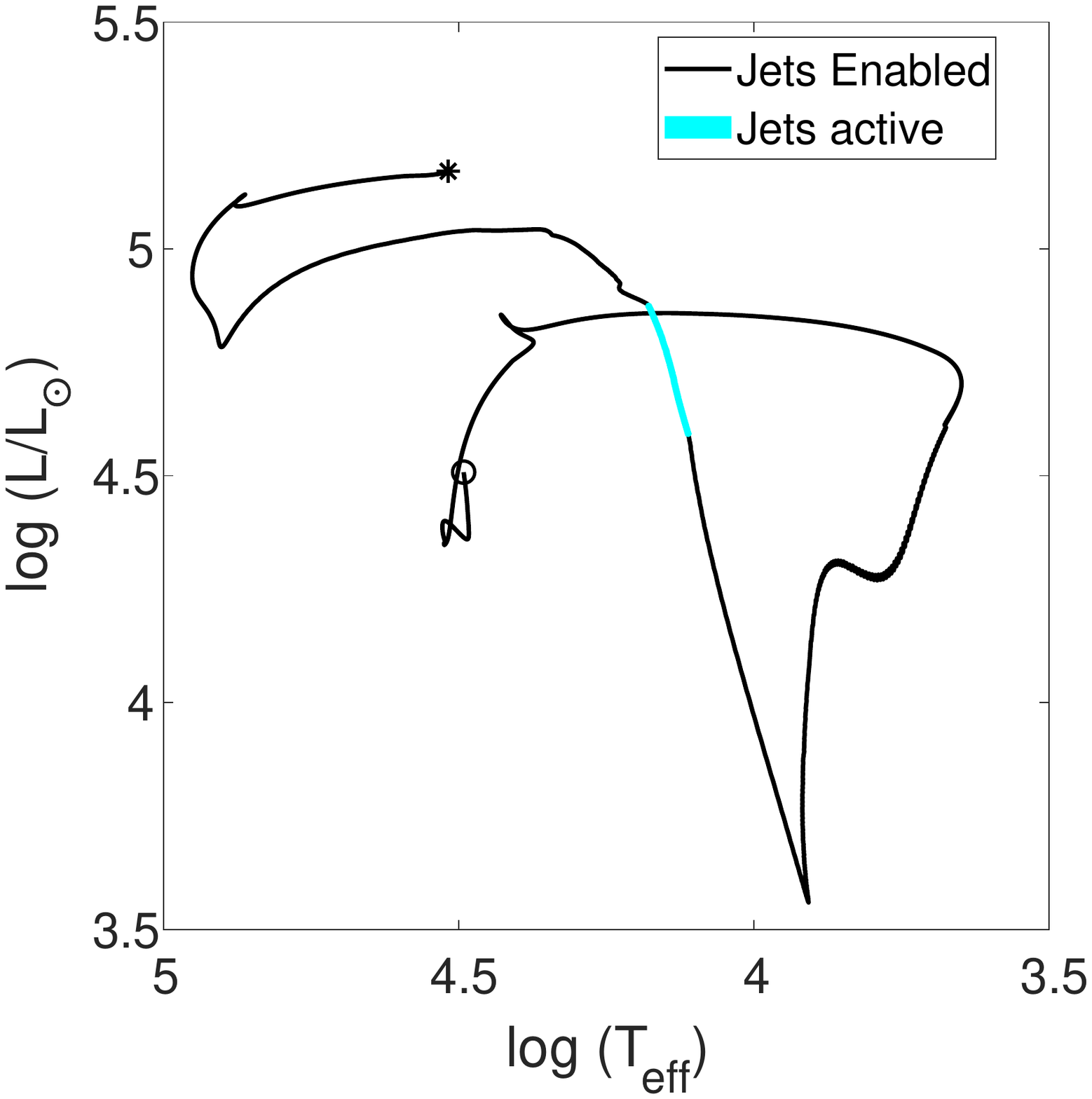}
\caption{The HR diagram of the primary star that we present in the left panels of figure \ref{fig:1000,e=0.4,fiducialded wind}, i.e., for the case with initial semi-major axis of $a_0=1000$ and initial eccentricity of $e_i=0.4$. The teal part of the line marks the duration for which jets were active. A circle marks the simulation starting point and an asterisk simulation termination as a CCSN.}
\label{fig:HRa1000e04}
\end{center}
\end{figure}

In Fig. \ref{fig:1000,e=0.4,fiducialded wind} we examine the evolution of a system with initial eccentricity of $e_i=0.4$, with, again, allowed jet activity (according to condition \ref{eq:fGEE}; left panels) and without jet activity (right panels). In this case, jets activity does take place, during the time period marked by the teal vertical band in the upper left panel. 
The primary reaches CCSN with a hydrogen mass of $M_{\rm H,f}=0.055 M_\odot$, hence a SN IIb. Without the enhanced mass loss rate due to jets activity, the system enters a CEE and does not get out from it. It is unlikely to form a SN IIb progenitor by this channel. This is a clear case where the GEE increases the parameter space for the formation of SN IIb progenitors. In Fig. \ref{fig:HRa1000e04} we present the HR diagram of the primary star in the case with jet activity (left panels of Fig. \ref{fig:1000,e=0.4,fiducialded wind}). 
\begin{figure*}
\begin{center}
 \includegraphics[trim=2cm 2cm 2cm 3cm,scale=0.65]{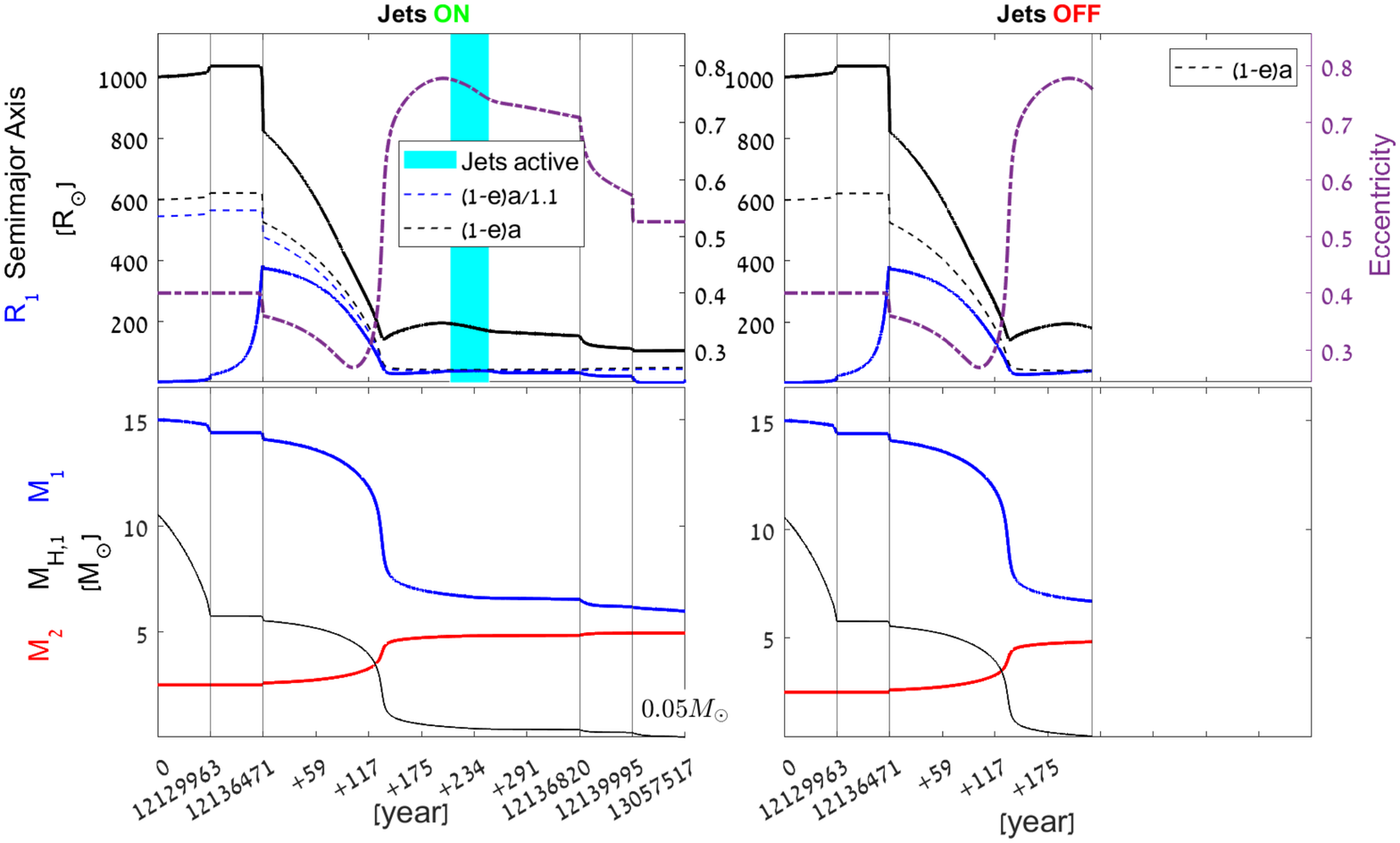}
\caption{Like Figure \ref{fig:1000,e=0.1,fiducialded wind}, but for an initial eccentricity of $e_i=0.4$. We terminate the evolution without jets activity (right panels) at the point where the binary system enters a CEE as it never recovered from it. This is an example of an eccentric system where the jet mechanism (a GEE) prevents a CEE. 
The five time periods on the left panel span, from left to right, $1.213 \times 10^7 \yr$ (earliest time segment), $6508 \yr$, $349 \yr$, $3175 \yr$, and $9.175 \times 10^5 \yr$ (last time segment).}
\label{fig:1000,e=0.4,fiducialded wind}
\end{center}
\end{figure*}

In Table \ref{Table1} we list simulations with other eccentricities, which we list in the first column, keeping all other initial parameters as in the other simulations of this section, i.e., $a_0=1000 R_\odot$,  $M_{\rm 1,i}=15 M_\odot$, $M_{\rm 2,i}=2.5 M_\odot$, and $f_{GEE}=1.1$. 
In the second column we indicate whether we allow (`On') or not (`Off') jets activity. 
The third column of Table \ref{Table1} lists the duration of the CEE, $\Delta t_{\rm CEE}$, where `Continues' implies that the system continues with the CEE and does not get out from it, and `No' indicates that no CEE occurred. The fourth column lists the final hydrogen mass in the primary star. 
Table \ref{Table1} shows that for a given initial semi-major axis, the probability of experiencing a CEE increases with increasing eccentricity. This is an expected result, as periastron distances are smaller for higher eccentricities. 
For $e_i \ga 0.4$ and the other parameters we use here, jets activity (the GEE) might play an important role. For other parameters jets might play an important role even from $e_i=0$ \citep{Naimanetal2020}. 
In some of the cases we present here jets prevent the CEE ($e_i=0$, $0.4$, $0.5$), in some cases it shorten the CEE phase ($e_i=0.6$, $0.8$), and in one case the jets enable the system to exit the CEE ($e_i=0.9$). 
One interesting case is $e_i=0.7$ where the CEE is longer when jets are active. This might result from a rapid mass removal by jets that causes the giant primary star to expand \citep{Naimanetal2020}. 
We emphasise again that we should treat with caution the evolution after a system enters a CEE, as the code does not treat well this phase (section \ref{subsec:Numerical}). 
In all cases where the system does not continue with the CEE the hydrogen mass fits a compact SN IIb progenitor.
\begin{table}[H]
\begin{footnotesize}
\begin{center}
\begin{tabular}{lccc}
Eccentricity      & Jets     & CEE duration   & Final $M_{\rm H}$ \\ 
   $e_i$               &          &    $\Delta t_{\rm CEE}(\yr)$       & $M_{\rm H,f}$($M_\odot$) \\
\hline 
0 (Fig. \ref{fig:1000,e=0,fiducialded wind})  & \color{red}Off  &  2.57  & 0.058    \\
0 (Fig. \ref{fig:1000,e=0,fiducialded wind})  & \color{green}On  & No  & 0.058   \\
0.1 (Fig. \ref{fig:1000,e=0.1,fiducialded wind})  & \color{red}Off  & No & 0.055   \\
0.1 (Fig. \ref{fig:1000,e=0.1,fiducialded wind})  & \color{green}On  & No    & 0.055 \\
0.2            & \color{red}Off  & No                    & 0.053        \\
0.2            & \color{green}On  & No                  & 0.047             \\
0.3            & \color{red}Off  & No                     & 0.052            \\
0.3            & \color{green}On  & No            & 0.046                          \\
0.4 (Fig. \ref{fig:1000,e=0.4,fiducialded wind})  & \color{red}Off&  Continues & $-$\\
0.4 (Fig. \ref{fig:1000,e=0.4,fiducialded wind})  & \color{green}On & No & 0.05  \\
0.5            & \color{red}Off  & 39.92              & 0.047          \\
0.5            & \color{green}On  & No                 & 0.047        \\
0.6            & \color{red}Off  &  111.77                    & 0.044                 \\   
0.6            & \color{green}On  & 39.20                & 0.044             \\
0.7            & \color{red}Off  &  190.05                    & 0.04              \\
0.7            & \color{green}On  & 268.18                  & 0.042               \\
0.8            & \color{red}Off  & 728.28                 & 0.039                \\
0.8            & \color{green}On  & 503.01         & 0.041               \\
0.9            & \color{red}Off  &  Continues    & $-$             \\
0.9            & \color{green}On  & 1474.04              & 0.04             \\
\hline
\end{tabular}
\caption{Twenty simulations with $a_0=1000 R_\odot$,  $M_{\rm 1,i}=15 M_\odot$, $M_{\rm 2,i}=2.5 M_\odot$, and $f_{GEE}=1.1$, but different initial eccentricities. If the system enters a CEE and does not get out, we terminate the evolution and cannot give the hydrogen mass at CCSN.}
\label{Table1}
\end{center}
\end{footnotesize}
\end{table}

Overall, in 2 out of the 10 case ($e_i= 0.4$, $0.5$) the jets activity through the GEE substantially helps the GEE channel for SNe IIb, and in few other cases it helps. Adding the results of \citep{Naimanetal2020}, we tentatively conclude that the GEE increases the number of systems that can form SNe IIb progenitors relative to the RLOF channel by tens of percents and possibly by a factor of two or somewhat more. 
   
\section{Other eccentric cases}
\label{sec:OtherCases}

In this section we present two more cases with initial semi-major axis of $a_0=600 R_\odot$ and $a_0=800 R_\odot$ and initial eccentricity of $e_i=0.5$. Additionally, we set $f_{\rm GEE}=1.2$ rather than $1.1$ as in previous runs. 
Figure \ref{fig:800,e=0.5,fiducialded wind} displays the case of $a_0=800 R_\odot$, and is another example of a case where jet activity (left panel) prevents the system from entering a continuous CEE (where the companion continues to spiral-in; right panel). 

\begin{figure*}
\begin{center}
 \includegraphics[trim=2cm 2cm 2cm 3cm,scale=0.65]{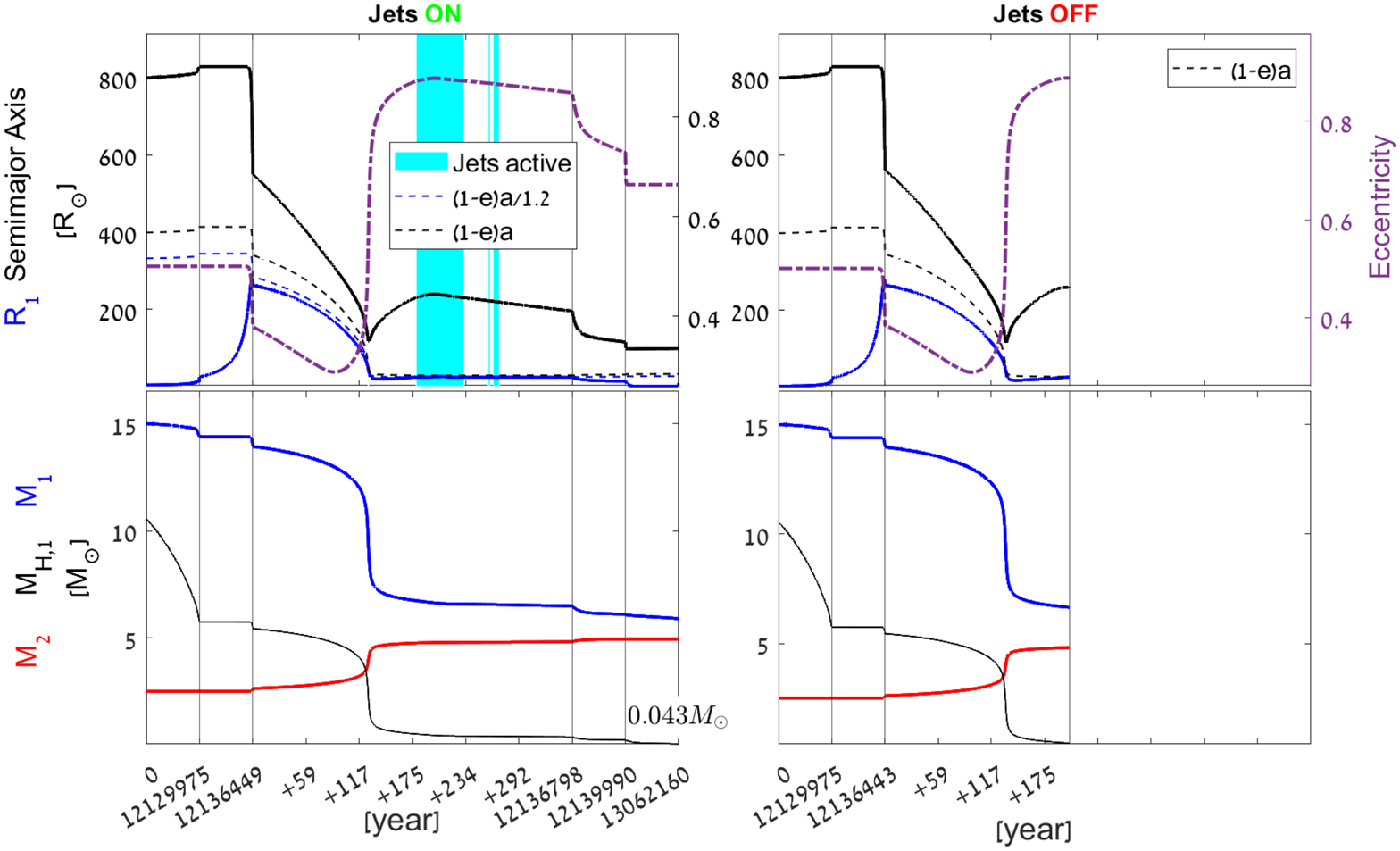}
\caption{Like Figure \ref{fig:1000,e=0.1,fiducialded wind}, but with an initial semi-major axis of $a_0=800 R_\odot$, an initial eccentricity of $e_i=0.5$, and $f_{\rm GEE}=1.2$ rather than $1.1$. The jet mass loss mechanism prevents CEE (left panel).
Without the extra mass loss due to jets the system enters a CEE and does not get out from it (right panel). The five time periods on the left panel span, from left to right, $1.213 \times 10^7 \yr$ (earliest time segment), $6474 \yr$, $349 \yr$, $3192 \yr$, and $9.222 \times 10^5 \yr$ (last time segment).}
\label{fig:800,e=0.5,fiducialded wind}
\end{center}
\end{figure*}
 
\begin{figure}[H]
\begin{center}
\includegraphics[trim=2cm 5cm 2cm 5cm,scale=0.4]{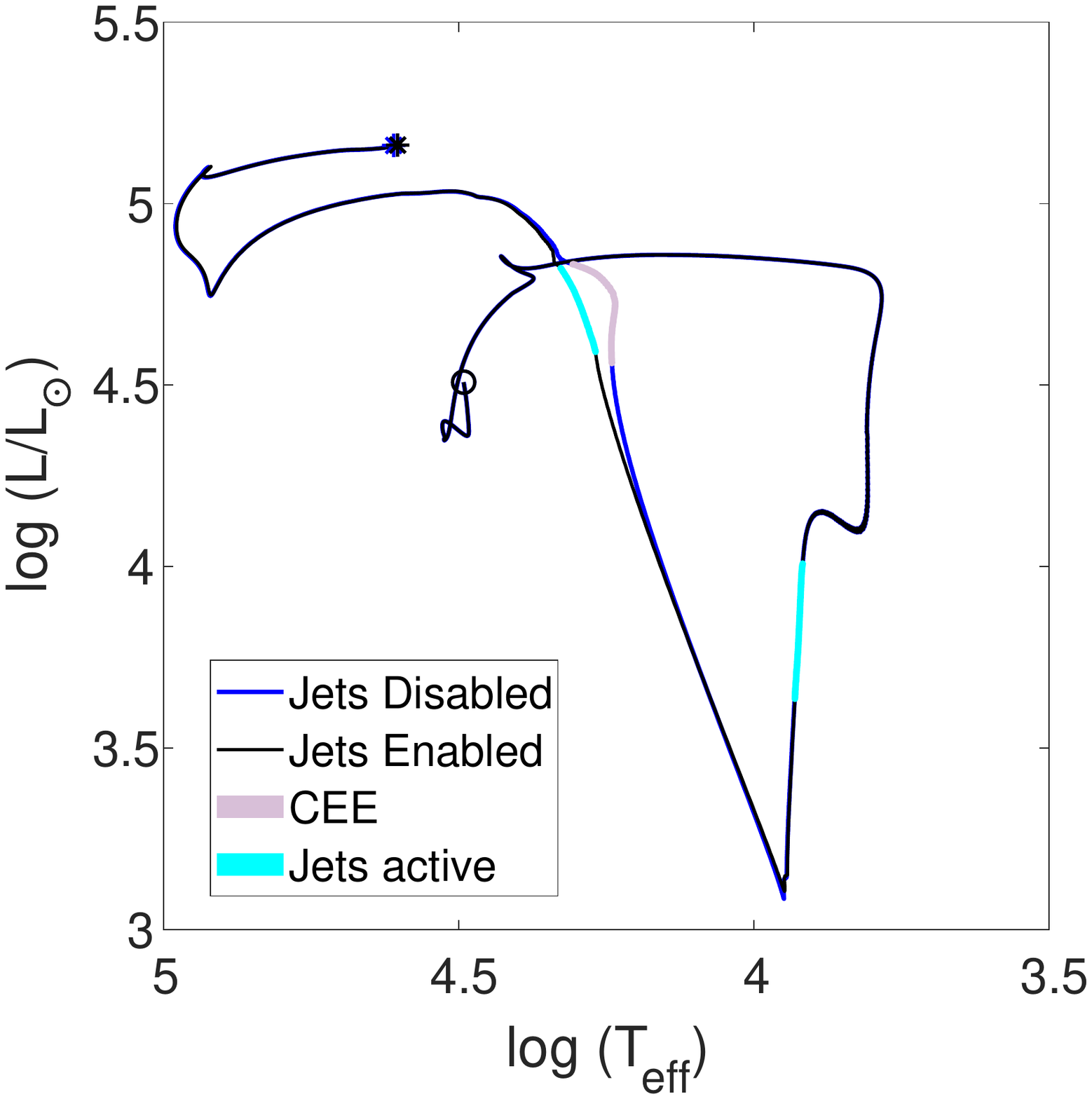}
\caption{The HR diagram of the primary stars for the cases with initial semi-major axis of $a_0=600$ and initial eccentricity of $e_i=0.5$ that we present in Fig. \ref{fig:600,e=0.5,fiducialded wind}. 
The black line denotes the case with jet activity (left panels of Fig. \ref{fig:600,e=0.5,fiducialded wind}), and the blue line denotes the case without jets activity (right panels of Fig. \ref{fig:600,e=0.5,fiducialded wind}). The teal parts mark periods of jets activity (as in the left panels of \ref{fig:600,e=0.5,fiducialded wind}), and the lilac parts denotes the CEE phase (as in the right panels of \ref{fig:600,e=0.5,fiducialded wind}). A circle marks the simulations starting point and an asterisk marks the termination of the simulations as CCSNe (the two cases end at about the same point on the HR diagram).}
\label{fig:HRa600e05}
\end{center}
\end{figure}

Figure \ref{fig:600,e=0.5,fiducialded wind} displays the case of $a_0=600 R_\odot$, and is another example of a case where jet activity prevents the system from entering CEE. 
We note that although the system without jets does exit the CEE phase, the CEE phase lasts for $\Delta t_{\rm CEE} =62 \yr \simeq 50 P_{\rm orb}$, where $P_{\rm orb}$ is the average orbital period during that phase. It is not clear that \textsc{mesa-binary} handles correctly such a long phase of a CEE. It might be that such a case will end in a continues CEE where the companion does spiral-in inside the envelope to a much smaller semi-major axis, and does not form a SN IIb progenitor by the GEE channel. In Fig. \ref{fig:HRa600e05} we present the HR diagram of the primary stars of the systems that we study in Fig. \ref{fig:600,e=0.5,fiducialded wind}. 
\begin{figure*}
\begin{center}
 \includegraphics[trim=2cm 2cm 2cm 3cm,scale=0.65]{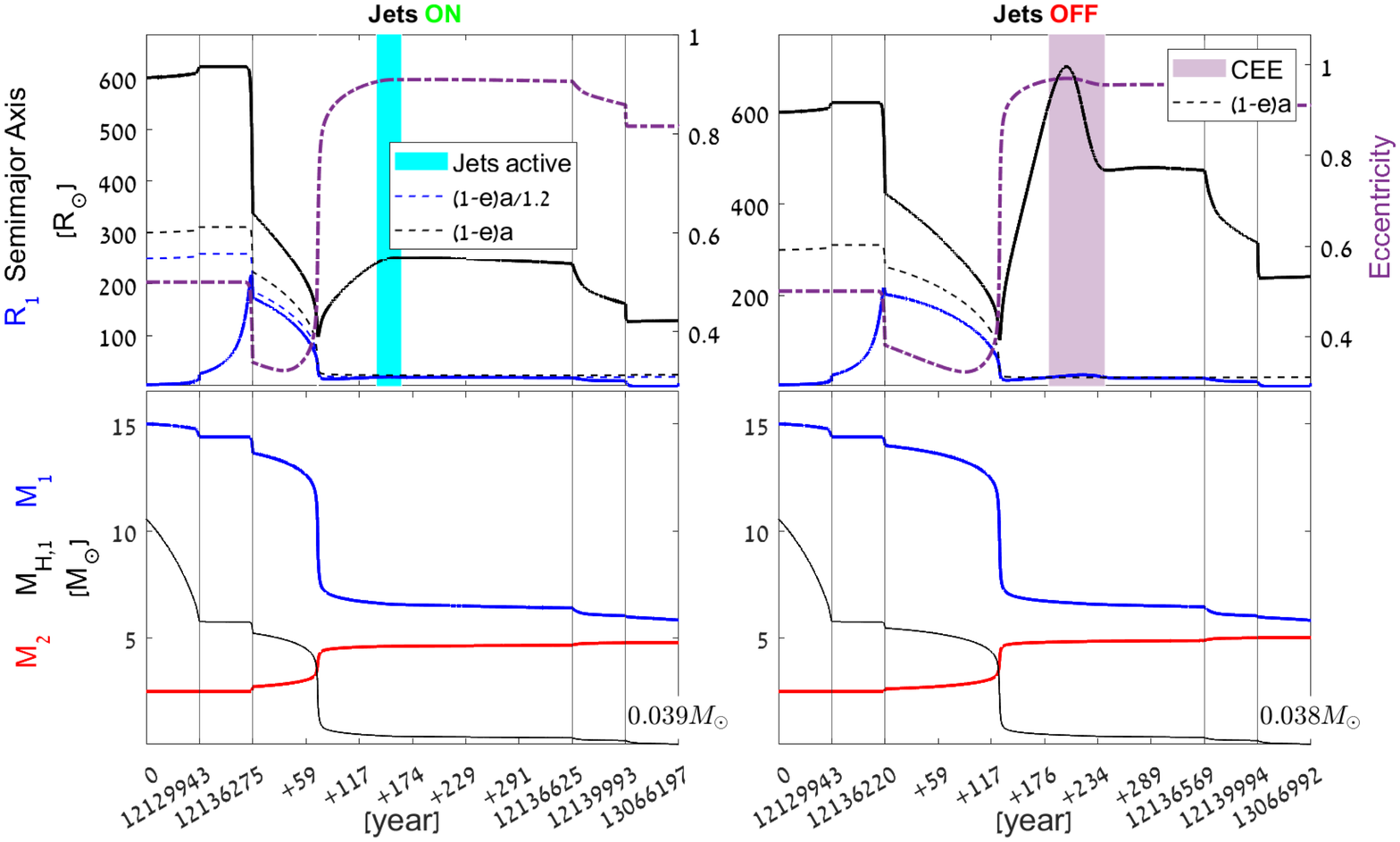}
\caption{Like Figure \ref{fig:800,e=0.5,fiducialded wind}, but with $a_0=600 R_\odot$. Note that the system enters CEE and later exits from the CEE. The five time periods span, from left to right, $1.213 \times 10^7 \yr$ (earliest time segment), $6332 / 6227 \yr$, $350 / 349 \yr$, $3368 / 3425 \yr$, and $9.262 / 9.304 \times 10^5 \yr$ (last time segment) on the left/right panels.}
\label{fig:600,e=0.5,fiducialded wind}
\end{center}
\end{figure*}

These two examples further demonstrate the role of jets in preventing the CEE in many cases, and by that in increasing the parameter space for the formation of SN IIb progenitors. We note again that not in all cases the jets are needed to prevent the CEE, and not in all cases the jets can prevent the CEE. 

Finally, we note that we conducted many more simulations, some of which we list in Appendix \ref{appensix:SimulationRuns}. We vary other parameters as we indicate in Table \ref{Table2}. 

\section{DISCUSSION AND SUMMARY}
\label{sec:summary}

This study deals with the GEE scenario for the formation of SN IIb progenitors. In this scenario the companion ends the evolution with a semi-major axis of $a_{\rm f} \approx 1 \AU$. By RLOF and the action of jets the secondary star removes a large fraction of the primary's envelope. Winds then remove more mass, leaving a small final hydrogen mass (Tables \ref{Table1} , \ref{Table2}). 
There is a class of low mass post-asymptotic giant branch (AGB) stars that have a main sequence companion at a semi-major axis of $\approx 1 \AU$, and which typically have eccentric orbits (e.g.,  \citealt{Kastneretal2010, VanWinckel2017}). In a large fraction of these the main sequence companion launches jets (e.g., \citealt{Thomasetal2013,  VanWinckel2017b, Bollenetal2019}). These post-AGB binary systems suggest that in most cases binary systems that experience the GEE end with eccentric orbits. This motivated us to extend an earlier study of the GEE scenario for SNe IIb \citep{Naimanetal2020} to include eccentric orbits. 

In that respect we note that the GEE predicts that the supernova and its circumbinary matter will have bipolar morphologies. Therefore, the GEE scenario is compatible with non-spherical explosion of SNe IIb, e.g., as Spectropolarimetric data analyses suggests for SN~1993J \citep{Stevanceetal2020}.
  
For that goal we used the stellar evolution code \textsc{mesa-binary} to simulate the evolution of eccentric binary systems, where we activate a jet-driven mechanism of enhanced mass loss. We compared each case with jets activity to a case without jet activity. We examined a variety of initial eccentricities and initial semi-major axes (section \ref{sec:eccentricity} and \ref{sec:OtherCases}; summarised in Table \ref{Table1}), as well as other values of several parameters (Appendix \ref{appensix:SimulationRuns}). We summarise our main results as follows.
\begin{enumerate}
  \item  Eccentric orbits (Figs. \ref{fig:1000,e=0.1,fiducialded wind} , \ref{fig:1000,e=0.4,fiducialded wind}) might have a different outcome than spherical orbits (Fig. \ref{fig:1000,e=0,fiducialded wind}) with the same semi-major axis. In particular, the final orbit is not circular, and might have very high eccentricity (e.g., Figs. \ref{fig:800,e=0.5,fiducialded wind},  \ref{fig:600,e=0.5,fiducialded wind}). 
    This is true with and without jets activity. 
  \item In cases with very low $e_i \simeq 0$, and moderate eccentricity, $0.4 \la e \la 0.5$, the jet-driven enhanced mass loss of the GEE can prevents a CEE (Fig. \ref{fig:1000,e=0.4,fiducialded wind}).
  This is also true for cases with an initial semi-major axis different from $a_0=1000$, as seen in Figures \ref{fig:800,e=0.5,fiducialded wind} and \ref{fig:600,e=0.5,fiducialded wind} (summarise in Table \ref{Table2}). The exact range of the eccentricities will change with other parameters. 
  \item For higher eccentricities the situation is more intricate. First, with and without jets the system might reach very high eccentricities of $e>0.99$, practically leading to merger. Only in some cases the jet-driven mass loss mechanism prevents this from taking place. On the other hand, in some cases the jets activity causes the system to enter a CEE. The reason is that rapid mass removal causes the giant envelope to expand more \citep{Naimanetal2020}.
  \item 
  In many cases the system enters a CEE and then gets out. As we discussed in section \ref{subsec:Numerical}, we do not expect  \textsc{mesa-binary} to treat well this phase, and uncertainties are large. In particular, if the CEE phase is long, we expect the system to continue to spiral-in to very small radii, rather than exit from the CEE. We found that in most cases when jets are active the CEE phase is shorter (Tables \ref{Table1} and \ref{Table2}), increasing the probability that the system will indeed exit the CEE and will form a SN IIb progenitor. 
\end{enumerate}

Overall, our results strengthen the conclusion of \cite{Naimanetal2020} that the process of jet-driven mass loss that leads to episode(s) of GEE phases substantially increases the binary parameter space that leads to the formation of SN~IIb progenitors. Like \cite{Naimanetal2020}, we also find that in all our cases the SN~IIb progenitors are blue-compact ones (rather than red supergiants progenitors). 
We do notice that, as other studies have shown (e.g., \citealt{Sravanetal2018, Naimanetal2020}), in some cases RLOF followed by a wind, even if the secondary star does not launch jets, can remove enough mass to form a SN~IIb progenitor. 
 
When we consider our results and those of \cite{Naimanetal2020}, we further strengthen the claim of \cite{Naimanetal2020} that the GEE increases the number of systems that can form SNe IIb progenitors relative to the RLOF channel by tens of percents and possibly by a factor of two or somewhat larger. 
 
Our results put the GEE channel on a more solid ground by including eccentric orbits, and do not change dramatically the conclusions of \cite{Naimanetal2020}. Namely, we do not change their estimate that $\approx 2-4 \%$ of CCSN progenitors experience the GEE under our assumptions, amounting to about quarter of all SNe IIb. The other three scenarios, each contribute about equal, are the binary evolution channel with RLOF but without GEE, the fatal-CEE, and the single-star channel (see list in section \ref{sec:intro}). 

\section*{Acknowledgments}

We thank an anonymous referee for useful comments and suggestions, and Avishai Gilkis for helpful discussions. 
This research was supported by a grant from the Israel Science Foundation and a grant from the Asher Space Research Fund at the Technion. We completed this work while the Technion was closed due to the Coronavirus (COVID-19). 

\textbf{Data availability}
The data underlying this article will be shared on reasonable request to the corresponding author.  

\appendix\section{More simulations}
\label{appensix:SimulationRuns}

In Table \ref{Table2} we list all our relevant simulations. 
We vary some parameters as we indicate in the different columns. Other parameters are as in the simulations we discuss in the main text, e.g., $M_{\rm 1,i}=15 M_\odot$ and $M_{\rm 2,i}=2.5 M_\odot$.

\nopagebreak[4]

\begin{table}[H]
\begin{center}
\begin{tabular}{lccccccc}
Semi-major Axis & $f_{\rm GEE}$ & $f_{\rm jets}$ & Eccentricity & Jets & CEE duration                & H mass \\
$a_0$           &               &                & $e_i$        &      & $\Delta t_{\rm CEE}$ ($\rm yr$) &  $M_{\rm H,f}$ ($M_\odot$) \\
\hline
1000 (Fig. \ref{fig:1000,e=0,fiducialded wind})& --- & - & 0   & \color{red}Off & 2.57  & 0.058 \\
1000 (Fig. \ref{fig:1000,e=0,fiducialded wind})& 1.1 & 3 & 0 & \color{green}On  & No  & 0.058   \\
1000 (Fig. \ref{fig:1000,e=0.1,fiducialded wind})& --- & - & 0.1 & \color{red}Off & No & 0.055  \\
1000 (Fig. \ref{fig:1000,e=0.1,fiducialded wind})& 1.1 & 3 & 0.1 & \color{green}On & No & 0.055 \\
1000 & --- & - & 0.2 & \color{red}Off  & No    & 0.053     \\
1000 & 1.1 & 3 & 0.2 & \color{green}On  & No   &  0.047      \\
1000 & --- & - & 0.3 & \color{red}Off  & No   &  0.052  \\
1000 & 1.1 & 3 & 0.3 & \color{green}On  & No    &  0.046  \\
1000 (Fig. \ref{fig:1000,e=0.4,fiducialded wind})& --- & - & 0.4 & \color{red}Off&Continues& $-$\\
1000 (Fig. \ref{fig:1000,e=0.4,fiducialded wind})& 1.1 & 3 & 0.4 & \color{green}On & No & 0.05 \\
1000 & --- & - & 0.5 & \color{red}Off  & 39.92     & 0.047 \\
1000 & 1.1 & 3 & 0.5 & \color{green}On  & No      &  0.047 \\
1000 & --- & - & 0.6 & \color{red}Off & 111.77 & 0.044 \\
1000 & 1.1 & 3 & 0.6 & \color{green}On & 39.20 &  0.044 \\
1000 & 1.2 & 4 & 0.6 & \color{green}On & No &  0.045 \\
1000 & --- & - & 0.7 & \color{red}Off & 190.05 *  & 0.04 \\
1000 & 1.1 & 3 & 0.7 & \color{green}On & 268.18 *  & 0.042 \\
1000 & 1.2 & 4 & 0.7 & \color{green}On & 114.06 * & 0.042 \\
1000 & --- & - & 0.8 & \color{red}Off & 728.28 * &  0.039 \\
1000 & 1.1 & 3 & 0.8 & \color{green}On & 503.01 *   & 0.041 \\
1000 & 1.2 & 4 & 0.8 & \color{green}On & 405.18 * & 0.042 \\
1000 & --- & - & 0.9 & \color{red}Off & Continues  & $-$ \\
1000 & 1.1 & 3 & 0.9 & \color{green}On & 1474.04 * & 0.04 \\
1000 & 1.2 & 4 & 0.9 & \color{green}On & 1353.13 *  & 0.04 \\

1000 & 1.2 & 3 & 0 & \color{green}On  & No & 0.058   \\
1000 & 1.2 & 3 & 0.1 & \color{green}On  & No & 0.055   \\
1000 & 1.2 & 3 & 0.2 & \color{green}On  & No & 0.053   \\
1000 & 1.2 & 3 & 0.3 & \color{green}On  & No & 0.051   \\
1000 & 1.2 & 3 & 0.4 & \color{green}On  & No & 0.049   \\
1000 & 1.2 & 3 & 0.5 & \color{green}On  & No & 0.047   \\
1000 & 1.2 & 3 & 0.6 & \color{green}On  & No & 0.045   \\
1000 & 1.2 & 3 & 0.7 & \color{green}On  & 123.82 & 0.042   \\
1000 & 1.2 & 3 & 0.8 & \color{green}On  & Continues & $-$   \\
800 & --- & - & 0 & \color{red}Off   & No  & 0.054    \\
800 & 1.2 & 3 & 0 & \color{green}On  & No & 0.054    \\
800 (Fig. \ref{fig:800,e=0.5,fiducialded wind}) & --- & - & 0.5 & \color{red}Off   & Continues & $-$  \\
800 (Fig. \ref{fig:800,e=0.5,fiducialded wind}) & 1.2 & 3 & 0.5 & \color{green}On  & No  & 0.043    \\
600 (Fig. \ref{fig:600,e=0.5,fiducialded wind})& --- & - & 0.5 & \color{red}Off & 61.93 & 0.038  \\
600 (Fig. \ref{fig:600,e=0.5,fiducialded wind}) & 1.2 & 3 & 0.5 & \color{green}On & No & 0.039    \\
\hline
\end{tabular}
\caption{A list of 39 simulations. 
The first four columns give some initial settings, while the fifth column indicates whether we allow for jets activity (the GEE; `On') or not (`Off'; see also explanation to Table \ref{Table1} in the main text). The sixth column lists the duration of the CEE if occurs, or indicates if no CEE occur (`No') or whether the system does not get out from the CEE and the secondary continues to spiral-in (`Continues'). 
The last column lists the final hydrogen mass in the primary star. 
If the system enters a CEE and does not get out, we terminate the evolution and cannot give the hydrogen mass at CCSN.
We also terminate evolution when the eccentricity becomes too large, $e>0.99$, or $e>0.9999$. 
We denote runs that had the eccentricity limit set to $e=0.9999$ rather than $e=0.99$ by a * in the "CEE duration" column.}
\label{Table2}
\end{center}
\end{table}

\label{lastpage}
\end{document}